\documentclass[prl,aps,amsmath,amssymb,amsfonts,floatfix,superscriptaddress,showpacs,twocolumn]{revtex4-1}
\usepackage{longtable}
\usepackage{graphicx}
\usepackage{epstopdf}
\usepackage{dcolumn}
\usepackage{epsfig}
\usepackage[dvips]{color}
\usepackage{dcolumn}
\usepackage{nicefrac}
\usepackage{float}
\usepackage{tabularx}
\newcolumntype{L}[1]{>{\raggedright\arraybackslash}p{#1}} 
\newcolumntype{C}[1]{>{\centering\arraybackslash}p{#1}} 
\newcolumntype{R}[1]{>{\raggedleft\arraybackslash}p{#1}} 

\newcommand{\bk}{\mathbf{k}}

\newcommand{\ham}{{\cal H}}

\def\beq{\begin{equation}}
\def\eeq{\end{equation}}
\def\be{\begin{equation}}
\def\ee{\end{equation}}

\newcommand{\cdaggop}[1]{c_{#1}^{\dagger}}

\def\t{\mbox{tr}\,}
\def\cG0{{\cal G}_0}
\def\cG{{\cal G}}

\def\a{\alpha}

\def\uc2{$U_{c2}$}
\def\uc1{$U_{c1}$}
\newcommand{\ket}[1]{\left| \left. #1 \right> \right.}

\def\bavs3{BaVS$_3$}

\def\t2g{$t_{2g}$}

\def\a1g{$a_{1g}$}

\usepackage[utf8]{inputenc}
\usepackage[T1]{fontenc}

\newcommand{\tdep}{\left(t\right)}

\begin{document}

\title{Extended dynamic Mott transition in the two-band Hubbard model out
of equilibrium}
\author{Malte Behrmann}
\affiliation{I. Institut f{\"u}r Theoretische Physik,
Universit{\"a}t Hamburg, 20355 Hamburg, Germany}
\affiliation{The Hamburg Centre for Ultrafast Imaging,
Luruper Chaussee 149,
22761 Hamburg, Germany}
\author{Michele Fabrizio}
\affiliation{International School for Advanced Studies (SISSA),
34136 Trieste, Italy}
\author{Frank Lechermann}
\affiliation{I. Institut f{\"u}r Theoretische Physik,
Universit{\"a}t Hamburg, 20355 Hamburg, Germany}

\begin{abstract}
We reformulate the time-dependent Gutzwiller approximation by M. Schir{\'o} and M. Fabrizio
[Phys. Rev. Lett. {\bf 105}, 076401 (2010)] in the framework of slave-boson mean-field theory, which is
used to investigate the dynamical Mott transition of the generic two-band Hubbard model at 
half filling upon an interaction quench. Interorbital fluctuations lead to notable changes  
with respect to the single-band case. The singular dynamical transition is 
replaced by a broad regime of long-lived fluctuations between metallic and insulating states, 
accompanied by intriguing precursor behavior. A mapping to a spin model proves helpful to 
analyze the different regions in terms of the evolution of an Ising-like order parameter. 
Contrary to the static case, singlet occupations remain vital in the Mott-insulating regime
with finite Hund's exchange.
\end{abstract}

\maketitle

\textit{Introduction}.--- 
Enormous advances in the physics of ultracold 
gases~\cite{bloch_many-body_2008,giorgini_theory_2008} have
evoked strong interest in time-dependent (TD) phenomena, which can be provoked 
and studied in trapped cold atom systems without all the complications that instead arise in solid-state materials.
This gives the unique opportunity to investigate physical realizations of prototypical models for interacting particles, such as Bose or Fermi Hubbard models,  and examine fundamental questions not only at equilibrium~\cite{jordens_mott_2008} but also in out-of-equilibrium conditions.~\cite{polkovnikov_colloquium:_2011,mark_preparation_2012,trotzky_probing_2012} The simplest protocol to drive a system out of equilibrium is 
a sudden change of its Hamiltonian parameters, for instance an interaction quench.  An intriguing issue  thereof is the possible trapping within 
metastable configurations that have no stable equilibrium counterpart.~\cite{eckstein_nonthermal_2008,polkovnikov_colloquium:_2011,werner_relaxation_2012,kollar_generalized_2011,carleo_localization_2012} Eventually, these metastable states decay at long times into some  thermal configuration; i.e., thermalization occurs. The detailed influence of thermalization onto TD
phenomena is a subject of its own.~\cite{eckstein_thermalization_2009,sciolla_quantum_2012,ates_thermalization_2012,canovi_quantum_2011,polkovnikov_colloquium:_2011} Considering the simple Fermi Hubbard model, most investigations 
concentrated so far on the single-orbital
case.~\cite{eckstein_thermalization_2009,schiro_time-dependent_2010} However,
materials in nature are most often ruled by multiorbital degrees of freedom. 
To what extent multiorbital processes modify the encountered single-orbital Hubbard physics 
out of equilibrium is a highly relevant question.

In this work we report on qualitative new physics close to the dynamic Mott transition of
the canonical two-band Hubbard model upon an interaction quench. The singular dynamic Mott 
transition in the single-band case is smeared out to a broad region with long-time fluctuations 
between metal and insulator. Unique precursor behavior takes place at the borders of this region.

\textit{Theoretical approach and model}.---
The dynamic Hubbard model is solved within TD slave-boson mean-field
theory (SBMFT) that merges the rotational invariant equilibrium slave-boson
method~\cite{li_spin-rotation-invariant_1989,lechermann_rotationally_2007,isidori_rotationally_2009} with a nonequilibrium scheme
recently proposed by Schir{\'o} and Fabrizio~\cite{schiro_time-dependent_2010,fabrizio_out--equilibrium_2012,schiro_quantum_2011} via the Gutzwiller
representation.~\footnote{A variant of the TD Gutzwiller technique was 
introduced~\cite{oelsen_time-dependent_2011} to compute response functions and has recently been 
extended~\cite{bunemann_linear-response_2013}.} In the latter scope the time evolution is 
described via coupled first-order differential equations for Slater determinants $\psi_0$ and 
Gutzwiller projectors $\phi$; i.e.,
\begin{eqnarray}
 \imath |\dot{\psi}_0\rangle &=& \tilde{H}[\phi] \ket{\psi_0}\;,\label{Gutz_DGL1}\\
 \imath \frac{\partial \phi} {\partial t} &=&\ham^{\rm loc}\,\phi +
\frac{\partial\,\langle\psi_0|\tilde{H}[\phi]|\psi_0\rangle}
{\partial \phi^\dagger}\;, \label{Gutz_DGL2}
\end{eqnarray}
whereby $\tilde{H}[\phi]$ denotes the renormalized free
Hamiltonian and $\ham^{\rm loc}$ the local interacting one (note that explicit site
dependance is omitted).
To reexpress these equations in SBMFT a unitary transformation of the Slater
determinants $\ket{\psi_0}$ into eigenstates $\nu^\bk$ of
$\tilde{H}[\phi]$ in momentum ($\bk$) space is performed. Hence the
former eigenstates are identified with the quasiparticle (QP) degrees of
freedom. For the complete transformation, utilizing $a,b$ for the eigenvalue labeling
and $A,B,C$ for the local basis states, one obtains
\begin{eqnarray}
\imath \frac{\partial \nu_{a\alpha}^\bk}{\partial t} &=&
\sum_\beta \tilde{H}_{\alpha \beta}^\bk
\nu_{a \beta}^\bk \label{dgl1_final}\;,\;\;\;
\tilde{H}_{\alpha \beta}^\bk =
\sum_{\alpha' \beta'} R^\dagger_{\alpha \alpha'}
 \varepsilon^{\bk}_{\alpha' \beta'} R_{\beta'\beta}\;,\quad\\
\imath \frac{\partial \phi_{AB}}{\partial t} &=& \sum_C \ham^{\rm loc}_{AC} \phi_{CB} +
\sum_{\bk b}^{\rm occ}\sum_{\alpha \beta} \nu_{b \alpha}^{*\bk}
\frac{\partial\tilde{H}_{\alpha \beta}^\bk} {\partial \phi_{AB}^\dagger}\nu_{b \beta}^\bk
\label{dgl2_final}\;.
\end{eqnarray}
with $\alpha,\beta$ denoting the respective orbital-spin combination.
The matrices $R_{\alpha\alpha'}$=$R_{\alpha\alpha'}[\phi]$ renormalize the free dispersion
$\varepsilon^{\bk}_{\alpha\beta}$ and form the QP-weight matrix
${\bf Z}$=${\bf R}{\bf R}^\dagger$.~\cite{lechermann_rotationally_2007}
Equations~(\ref{dgl1_final}) and (\ref{dgl2_final}) show the coupled time evolution of QP 
states $\nu^\bk$ and local slave-boson amplitudes $\phi_{AB}$. Within mean-field
versions of Gutzwiller and slave-boson techniques, those quantities are subject
to certain constraints, reading (for each point in time $t$)
\begin{equation}
\operatorname{Tr} \left(\phi^\dagger \phi  \right)=1 \;,\quad
\operatorname{Tr} \left(\phi^\dagger  \phi\,  \cdaggop{\alpha} c_\beta \right)
=\sum_{\bk b}^{occ} \nu_{b \alpha}^{*\bk} \nu_{b \beta}^\bk \quad.
\label{SBMFT_constraint}
\end{equation}
A numerical solution of Eqs.~(\ref{dgl1_final}) and (\ref{dgl2_final}) is achieved by using 
an adaptive Runge-Kutta scheme of order 5/6.~\cite{verner_numerically_2010}

\begin{figure}[t]
\centering
\includegraphics*[width=8.3cm]{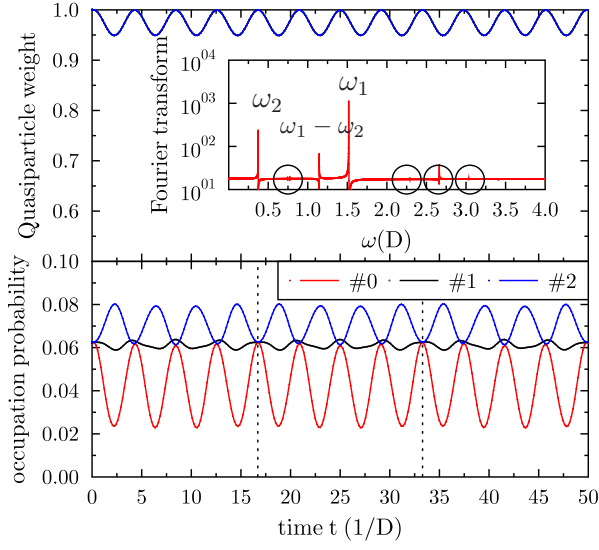}%
\caption{(Color online) QP weight $Z(t)$ and occupation probability \#$p$ of local
sectors with particles $p$=0,1,2 for the interaction quench to $U_f$=0.2. Dotted lines denote
period of local states oscillations. The inset shows the Fourier transform (absolute value of Fourier coefficients in logarithmic
units) of $Z(t)$. See the text for use of the circles.}
\label{fig:example}
\end{figure}
We study the canonical two-band Hubbard Hamiltonian $\ham=H+\sum_i\ham_i^{\rm loc}$ 
with a nearest-neighbor hopping $\tau$ that defines the kinetic part $H$. In detail it reads
\begin{eqnarray}
\ham&=& -\tau \sum_{\langle i,j\rangle m\sigma}
\left(\cdaggop{im\sigma} c^{\hfill}_{jm\sigma}+{\rm h. c.}\right)
+U\sum_{im} n_{im\uparrow}n_{im\downarrow}+\nonumber\\
&&\hspace*{-0.5cm}+\frac 12 \sum \limits _{i,m \ne m',\sigma}
\Big\{U' \, n_{im \sigma} n_{im' \bar \sigma}
+ U'' \,n_{im \sigma}n_{im' \sigma}+\\
&&\hspace*{-0.5cm}+\left.J\left(c^\dagger_{im \sigma} c^\dagger_{im' \bar\sigma}
c^{\hfill}_{im \bar \sigma} c^{\hfill}_{im' \sigma}
+c^\dagger_{im \sigma} c^\dagger_{im \bar \sigma}
 c^{\hfill}_{im' \bar \sigma} c^{\hfill}_{im' \sigma}\right)\right\}\;,\nonumber
\label{eq:hubbardham}
\end{eqnarray}
where $i,j$ are site indices, $m$,$m'$ run over orbitals 1, 2, and
$\sigma$=$\uparrow,\downarrow$ marks the spin projection, i.e., $\alpha, \beta = m\sigma$,$m'\sigma$ in connection with (\ref{dgl1_final}) and (\ref{dgl2_final}). A three-dimensional simple-cubic
dispersion is used and thus the parametrization $U'$=$U$$-$$2J$, $U''$=$U$$-$$3J$ proves
adequate.~\cite{castellani_magnetic_1978,fresard_interplay_1997} The value of the hopping $\tau$ is 
such that the half-bandwidth $D$ is the energy unit. 
To investigate an interaction quench the initial $U_i$ is set to zero and the local 
interaction varies in time as $U(t)$=$U_f\,\Theta(t)$ and $J(t)$=$qU_f\,\Theta(t)$ , i.e., 
jumps from zero to $U_f$ and $qU_f$ at $t$=0. In the following we focus on the paramagnetic
half-filled scenario and aim at general dynamic multiorbital Mott-transition mechanisms; thus antiferromagnetic fluctuations on the specifically chosen lattice type
are neglected.
\begin{figure}[b]
\centering
\includegraphics*[width=8.3cm]{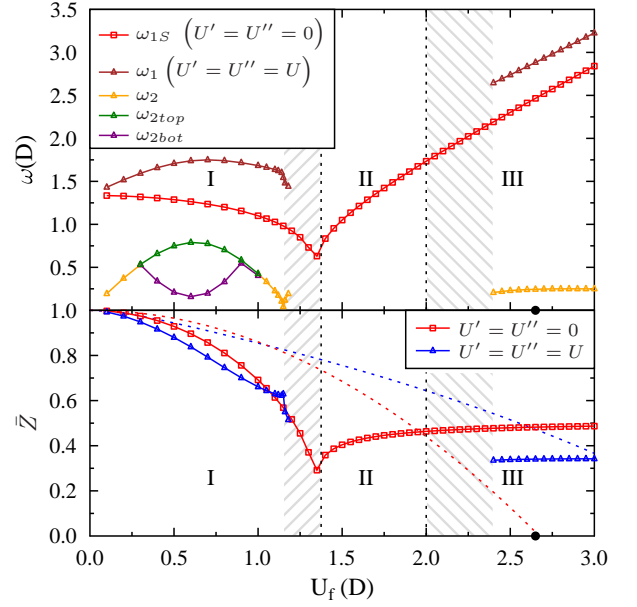}%
\caption{(Color online) Top: Main-frequency evolution with $U_f$ for $U'$=$U''$=0 and
$U'$=$U''$=$U$ from Fourier transforming $Z(t)$. The dotted lines separate three
different regions (see text and Fig.~\ref{fig:order_parameter_so2onlyU}).
Bottom: Time-averaged $Z$ as a function of $U_f$. 
Dashed lines mark the equilibrium behavior with the black dot denoting the associated Mott-critical
point. Gray hashed area indicates precursor behavior.}
\label{fig:zmat_freq}
\end{figure}

\textit{Results}.---
First the limiting case $J$=0 ($q$=0), i.e., $U$=$U'$=$U''$, is examined.
The quench leads to nonvanishing oscillations in the physical observables due to the
lack of quantum fluctuations in the present formalism.~\cite{schiro_time-dependent_2010}
In Fig.~\ref{fig:example} the TD QP weight
$Z_{m \sigma}\tdep$=$\delta_{12} \delta_{\uparrow \downarrow} Z \tdep$ and
occupation probability \#$p$ of local particle sectors is displayed for small $U_f$.
While $Z\tdep$ and the occupation in the $p$=0,1 sectors oscillate in phase, the
two-particle sector commutes with a $\pi$ phase shift thereto. As expected, minima in
$Z\tdep$ amount to maxima in \#2 and vice versa. The Fourier transform (inset in
Fig.~\ref{fig:example}) reveals the occurrence of two main frequencies $\omega_{1}$,
$\omega_{2}$ as well as sidebands. This is in contrast to the single-band and the
quasi-decoupled ($U'$=$U''$=0) two-band case, where only a single frequency on the order of
$\omega_{1}$ shows up. While the latter frequency in the fully interacting case is 
sensitive to changes in the hopping $\tau$, $\omega_{2}$ remains rather unaffected. 
Hence $\omega_{2}$ originates from interorbital processes and is absent for the 
quasi-decoupled-band case.
The encircled frequencies are integer multiples of $\omega_{1}$, $\omega_{2}$ or
sidebands and appear due to Fourier transformation on a finite-time 
interval (note logarithmic units). The time interval was set to 7500 $D^{-1}$ to provide 
suitable high-frequency resolution $<$$(1000D)^{-1}$.
\begin{figure}[t]
\includegraphics*[width=7cm]{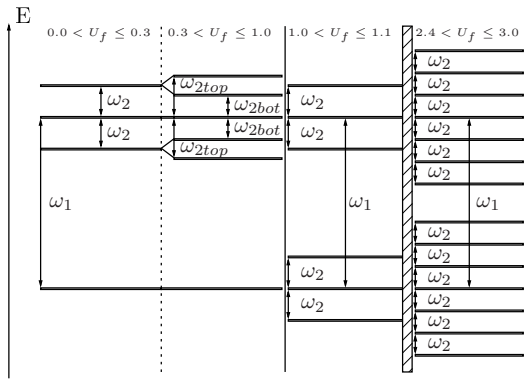}%
\caption{Level diagram for the active frequencies in the case $U'$=$U''$=$U$ 
within the different regimes dictated by $U_f$.}
\label{fig:energy-lev}
\end{figure}

The overall relevant-frequency behavior from the Fourier transform of $Z(t)$ with
increasing $U_f$ is depicted in Fig.~\ref{fig:zmat_freq} with comparison to the
$U'$=$U''$=0 case.
For the latter, the bands are independent and we recover the results of the 
single-band study.~\cite{schiro_time-dependent_2010} A dynamic Mott transition occurs
at $U^{(1)}_{c}$=1.325 eV indicated by a logarithmic divergence in the single
frequency $\omega_{1S}$, smeared out due to the numerical integration. With
interorbital terms, multiple relevant frequencies appear with growing
$U_f$. In fact, $\omega_2$ consists of two coupled interorbital
contributions, namely $\omega_{2top}$ and $\omega_{2bot}$. Both contributions tend to
zero close to the dynamic Mott criticality, which we believe to occur near $U^{(1)}_{c}$
(see below). 
Three observations from Fig.~\ref{fig:zmat_freq} are vital. The first is the splitting of $\omega_2$ for
0.3$<$$U_f$$<$1.0 which can be visualized within a level scheme
(see Fig.~\ref{fig:energy-lev}). That scheme shows the onset of splitting between levels 
connected by $\omega_{1}$ when approaching the Mott critical regime. The second is that a broad region of 
mostly noisy Fourier spectra appears for 1.15$<$$U_f$$<$2.375 without a clear structuring. 
Albeit already insulating, a definite level structure then reemerges for $U_f$$>$2.375.

Whereas $\omega_2$ saturates for large interaction quenches, $\omega_1$ linearly rises with 
the same slope as in the quasi-decoupled-bands case. Figure~\ref{fig:zmat_freq} also shows the
quantity $\bar{Z}$=$\bar{Z}(T)$$\equiv$$\frac{1}{T} \int_{0}^{T}\hspace{-0.1cm}dt\,Z$,
based on time-averaging over a long-time interval $T$ (not to be confused with a basic period
of oscillation), in comparison with the equilibrium QP weight. If not specified $T$ was set to 7500 $D^{-1}$. A well-defined 
$\bar{Z}$ is only accessible outside the broad $U_f$ range with noisy Fourier spectrum.
Interestingly, just before entering the latter regime a small range appears where
$\bar{Z}$ of the $U'$=$U''$=0 model is {\sl smaller} than in the one with including 
interorbital terms. Thus interorbital interactions allow one to {\sl reduce} the 
standard correlation measure in certain out-of-equilibrium cases.

Better understanding of the dynamical transition at $J$=0 can be gained by an Ising-spin 
representation of the model that is fully equivalent to slave-bosons and 
reads~\cite{fabrizio_out--equilibrium_2012}
\begin{equation}
\ham_S = -\frac{J}{S^2}\frac{2}{r} \sum_{\langle i,j \rangle} S_{ix} S_{jx} +
\frac{U}{2} \sum_i \left(S_{iz}\right)^2\quad,
\label{eq:spinham}
\end{equation}
with spin $S$=2 and where $r$=6 is the lattice coordination number while $-J$=$-2/3$ is the 
energy per site of the non interacting ground state. Within mean field, i.e., assuming a 
variational wave function $\prod_i\, |\Phi(S_{iz})\rangle$, 
the spin model above becomes identical to the slave-boson mean-field theory at half filling 
if  $|\Phi(S_{iz})\rangle = \sum_{p=0}^4\, \phi_{p}\,| S_{iz}=p-2\rangle$, where $\phi_{p}$ 
is the original slave boson at site in particle sector $p$. Because of half filling, 
$\phi_{p}$=$\phi_{4-p}$, so that $\langle S_{iz}\rangle$=0. 
Metallic coherence is signaled in the spin model by a finite Ising order parameter 
$\langle S_{ix}\rangle$=$\Re\Big[\sqrt{6}\big(
\phi_3^*\,\phi_2 + \phi_2^*\phi_1\big) + 2\big(\phi_4^*\phi_3 + 
\phi_1^*\phi_0\big)\Big] \equiv M_1 + M_2$, 
while the incoherent Mott insulator has $\langle S_{ix}\rangle$=$\langle S_{iz}^2\rangle$=0. 
The initial noninteracting state is characterized by $\langle S_{ix}\rangle$=2 and 
$\langle S_{i z}^2\rangle$=1, i.e. energy per site $E$=$-2/3$$+$$U/2$, conserved during the 
unitary evolution. If, like in the single-band case, \cite{schiro_time-dependent_2010} we 
assume the dynamical Mott transition to occur when the energy equals that of the Mott 
insulator, i.e., $E$=0, then we would expect $U^{(2)}_c$=4/3, which is also equal to the 
single-band value for the dynamical Mott transition. In fact, this argument predicts one 
and the same value $U_c$=4/3 for any $N$-band simple, i.e., $J$=0, Hubbard model at 
half filling, indeed close to the numerical value.
\begin{figure}[t]
\centering
\includegraphics*[width=8.5cm]{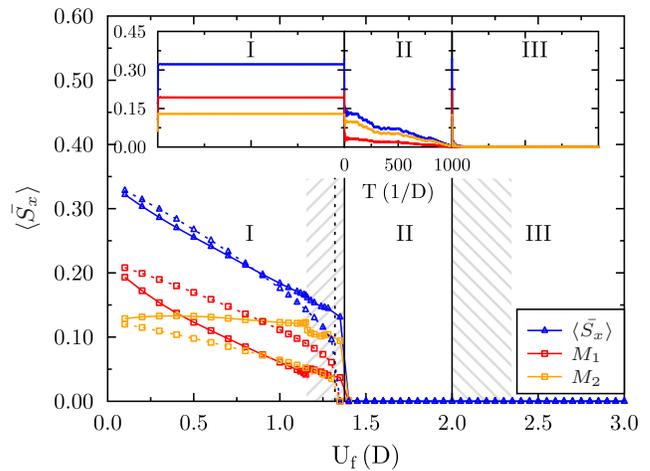}%
\caption{(Color online) Order parameter $\bar{\langle S_x \rangle}$
for $U'$=$U''$=0 (dashed) and $U'$=$U''$=U (full). Inset shows
$\bar{\langle S_x \rangle} \left( T \right)$ in the three different regions. The
quasidecoupled two-band case stabilizes only in regions I and III 
(divided by the vertical dashed line). Region II appears with interorbital terms.
The precursor regime is visualized by the gray hashed area.}
\label{fig:order_parameter_so2onlyU}
\end{figure}

Figure~\ref{fig:order_parameter_so2onlyU} exhibits the time-averaged quantity
$\bar{\langle S_x \rangle}$ with increasing $U_f$, indeed classifying three different regions. 
Hence $\bar{\langle S_x \rangle}$ does serve here as an order parameter, where regime I is metallic 
with a finite $\bar{\langle S_x \rangle}$. However there is a crossover of the
contributions $\bar{M}_1$, $\bar{M}_2$ for $U'$=$U''$=$U$, showing that correlations
between local states of the one- (three-) and two-particle sectors are less affected by
increasing $U_f$ than those between zero- (four-) and one- (three-) particle sectors. 
For $U'$=$U''$=0 the order parameter jumps to zero at $U_c^{(1)}$, signaling the dynamic 
metal-insulator transition into a regime III. With finite interorbital interactions the 
behavior is surprisingly more intriguing. At $U_f=$1.15 a precursor regime starts with 
increased fluctuations of growing frequency. The integrated components $M_1$, $M_2$ are 
specifically sensitive thereto. This may be connected to the breakdown of $\omega_2$
shown in Fig. \ref{fig:zmat_freq}, leading to a noncontinuous evolution of correlations
between the local particle sectors. In the region 1.375$\le$$U_f$$\le$2.0 the order 
parameter then indeed vanishes, but only after an extremely long time. This extended 
Mott-insulating transition (EDMT) is characterized by a chaotic-like time evolution of the 
QP weight and the multiplet-occupation probabilities. This is in accordance with the
noisy Fourier spectra. Note that the EDMT is a unique feature of the general two-band model with 
interorbital interactions. The static Mott-critical interaction strongly differs if the orbitals 
are explicitly coupled ($U_c$$\sim$2.65) or not ($U_c$$\sim$4.0). In contrast the boundary of 
the dynamic metallic region I is quite independent of the coupling type, but a qualitative 
difference is introduced via the appearance of an intermediate region for interaction-coupled 
orbitals.
\begin{figure}[t]
\centering
\includegraphics*[width=8.3cm]{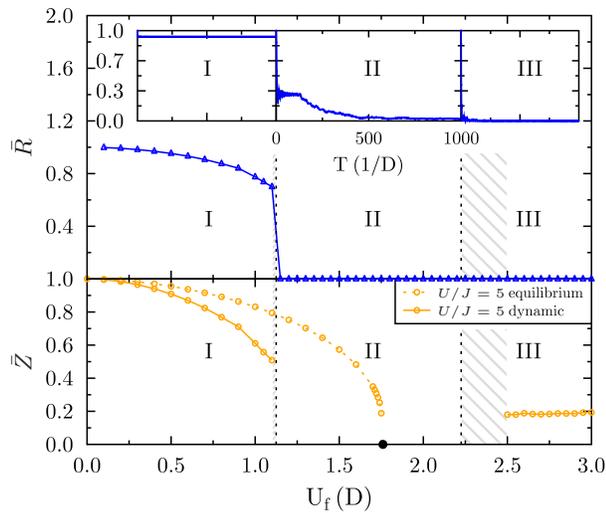}%
\caption{(Color online) Time-averaged diagonal renormalization $R$ and QP weight $Z$
for $U/J=5$. Black dot marks the Mott-critical point of the associated equilibrium model. Insets show $\bar{R}(T)$ in the three different regions.}
\label{fig:rmat_j5}
\centering
\end{figure}

\begin{figure}[t]
\centering
\includegraphics*[width=8.3cm]{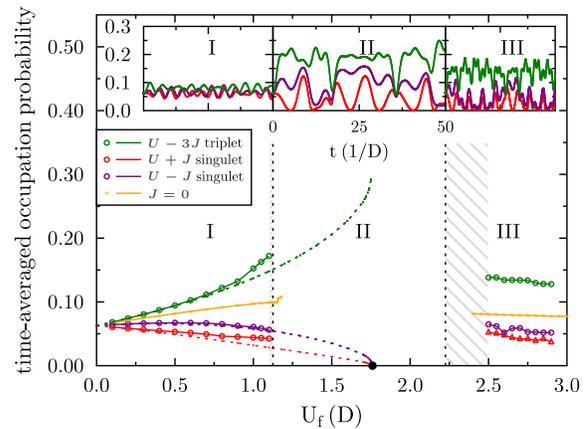}%
\caption{(Color online) Time-averaged multiplet occupation probabilities for $U/J=5$ with
the equilibrium probabilities (dashed lines) for comparison. The Mott-critical point of the associated equilibrium model is described by a black dot.
Insets reveal explicit time evolution of multiplet occupation probabilities in different regions.}
\label{fig:phi_j5}
\end{figure}

The EDMT is not an artifact of the ($J$=0) case but a general feature of the
general dynamic two-band Hubbard model. Since $\bar{\langle S_x \rangle}$ is proportional
to the $R$ matrix [see Eq.~(\ref{dgl2_final})], the quantity $\bar{R}$ serves as a
suitable order parameter in the original model and is depicted for $U/J$=$1/q$=5
[including all interaction terms in Eq.~(6)] in  Fig.~\ref{fig:rmat_j5}.
The precursor regimes as well as the occurrence of the EDMT persists. In fact with
finite $J$ this intermediate region is even broadened (1.125$\le$$U_f$$\le$2.225) 
in the $\bar{R}$-$U_f$ diagram compared to the $J$=0 case (1.375$\le$$U_f$$\le$2.0). In addition, 
Fig.~\ref{fig:phi_j5} shows the evolution of the time-averaged multiplet-occupation 
probabilities for finite $J$. For small $U_f$ they are close to the values in the 
equilibrium model, but lack the strong polarization in the insulating region. While in 
the static case only the triplet channels survive the Mott transition, the dynamic Mott 
region III displays also finite singlet fillings.

\indent \textit{Summary\label{sec:dis}}.---
We have investigated an interaction quench in the canonical two-band Hubbard model with
the TD-SBMFT scheme allowing for complete rotational invariance. The limited case
with sole intraorbital interaction terms leads to a quasi decoupled two-band model with
dynamic characteristics reminiscent of prior single-band
studies.~\cite{schiro_time-dependent_2010} On the other hand, when introducing the
relevant interorbital interactions, novel physics appears out of equilibrium. An
intermediate region with long-time chaotic-like fluctuations in the physical amplitudes 
and high-frequency metal-to-insulator fluctuations emerges and replaces the singular 
Mott-transition point. This replacement goes along with precursor regimes exhibiting 
high-frequency fluctuations in time. Finally the appearance of chaotic-like behavior may be 
connected to the non integrable classical coupled-pendulum problem, which indeed 
displays chaotic orbits. Extensions beyond mean field are needed to reveal whether such analogies
hold in the complete quantum-fluctuating scenario.

\begin{acknowledgments}
This work has been supported by the DFG cluster of excellence ``The Hamburg Centre
for Ultrafast Imaging'' as well as the  DFG-SFB925 and by EU-FP7 under the project GO FAST No. 280555. 
Computations were performed at the North-German Supercomputing
Alliance (HLRN) under Grant No. hhp00026.
\end{acknowledgments}
\bibliographystyle{apsrev4-1}
\bibliography{Paper1}
\end{document}